\documentclass[a4paper, 11pt]{article}

\usepackage[utf8]{inputenc}
\usepackage{authblk}
\usepackage[english]{babel}
\usepackage{times}
\usepackage{hyperref}
\usepackage{graphicx}
\usepackage{parskip}

\usepackage[font=small,labelfont=bf]{caption}

\title{The Westermo test results data set}

\author[*]{Per Erik Strandberg}

\affil[*]{Westermo Network Technologies AB, Västerås, Sweden%
\authorcr Email: per.strandberg@westermo.com}

\date{November 2022 (version 1.0)}

\begin{document}

\maketitle

\section*{Abstract}
There is a growing body of knowledge in the computer science, software engineering, software testing and software test automation disciplines. However, there is a challenge for researchers to evaluate their research findings, innovations and tools due to lack of realistic data. 
This paper presents the Westermo test results data set, more than one million verdicts from testing of embedded systems, from more than five hundred consecutive days of nightly testing. The data also contains information on code changes in both the software under test and the test framework used for testing.
This data set can support the research community in particular with respect to the regression test selection problem, flaky tests, test results visualization, etc.

\phantom{fake paragraph for space}

\noindent
\textbf{Keywords:}
software engineering, software testing, embedded systems, cyber-physical systems, nightly testing, regression testing, test case prioritization, flaky tests, test results visualization.

\section{Specifications}

\begin{description}    
\item[Subject:]{Computer Science -- Embedded Systems, and Software Engineering}
\item[Specific subject area:]{Nightly testing of software in networked embedded systems.}
\item[Type of data:]{Table}
\item[How the data were acquired:]{Test results data were acquired using Westermo's test automation framework in the Westermo test software test lab. Data from source code changes in the software under test and in the test framework was stored in the git source code management system. See related research article for more details.}
\item[Data format:]{Raw}
\item[Description of data collection:]{Test results data were exported from the Westermo test results database, and anonymized. Data from source code changes were exported from Westermo's source code repository, and anonymized.}
\item[Data source location:]{Data was collected at Westermo Network Technologies AB, on Metallverksgatan, in Västerås, Sweden.}
\item[Data accessibility:]{Data is available at GitHub:
\href{https://github.com/westermo/test-results-dataset}{https://github.com/westermo/test-results-dataset}
}
\item[Related research article:]{
P.\ E.\ Strandberg,
\emph{Automated System-Level Software Testing of Industrial Networked Embedded Systems,}
PhD Thesis, Mälardalen University, 2021.
ISBN: 978-91-7485-529-6.

Archived in arXiv: \href{https://arxiv.org/abs/2111.08312}{https://arxiv.org/abs/2111.08312} as well as in the Swedish National Library \href{http://urn.kb.se/resolve?urn=urn:nbn:se:mdh:diva-56036}{http://urn.kb.se/resolve?urn=urn:nbn:se:mdh:diva-56036}. 
}
\end{description}

\section{Value of the data}

\begin{figure}[t]
  \centering
    \fbox{\includegraphics[width=0.65\linewidth]{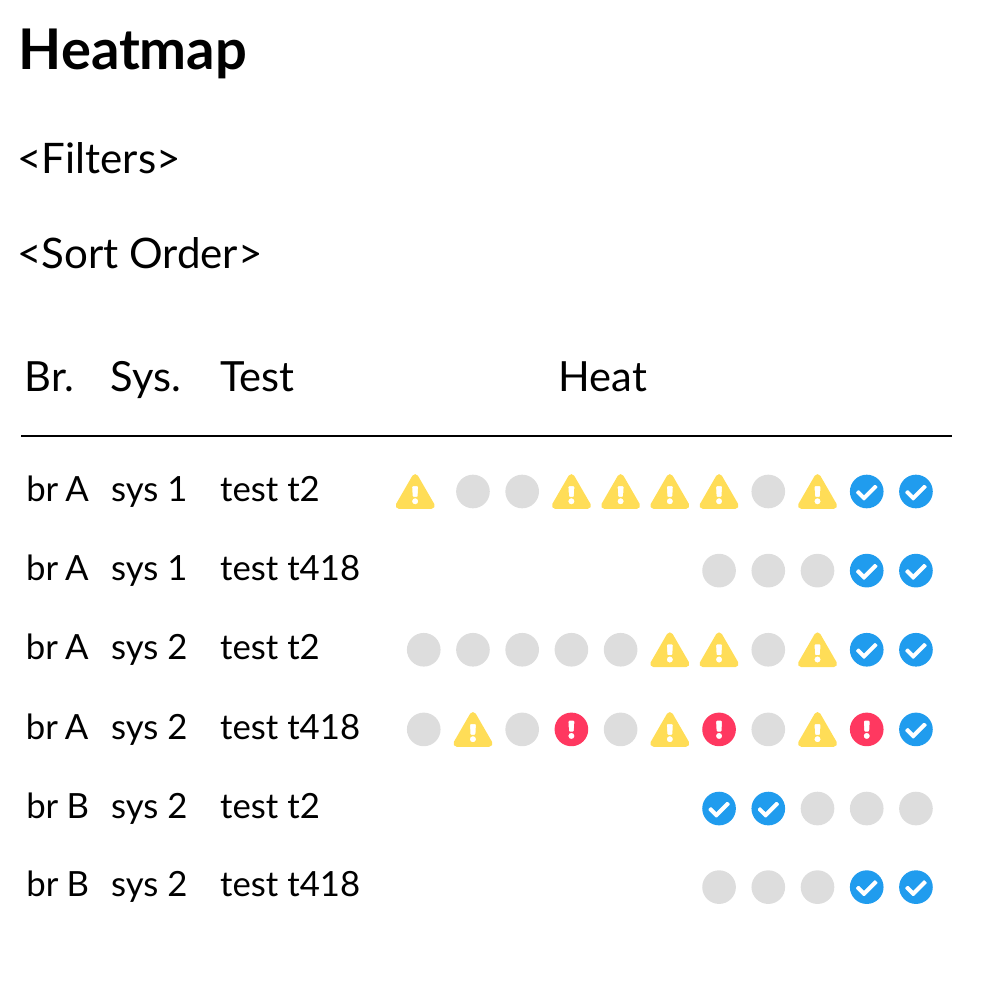}}
  \caption{
  The ``heatmap'' test results visualization approach that shows results from several test cases, being tested on multiple code branches, on several test systems, over time.
  The icons represent different verdicts:
  blue means pass (0),
  red means fail (1),
  yellow means invalid (2), and the gray icons 
  represent the skipped verdict -- there was no time to run this test -- which was not in use
  at Westermo at the time of data collection. 
  Note that the figure uses visual redundancy in that both color and symbol is used to represent values, and that the combination red and green is avoided to simplify understanding for color-blind viewers.
  Test t418 is flaky for branch A on test system 2.  
  Figure originally from \cite{strandberg2022software}.
  }
  \label{fig:tim-visualization}
\end{figure}

This data can be useful for researchers working on test automation, e.g.\ we have used it to support work on regression test selection \cite{felding2022thesis,strandberg2016,strandberg-nutshell},
flaky tests \cite{strandberg2020intermittently},
test results visualization \cite{strandberg-heatmaps, strandberg2022software} (see example in Figure~\ref{fig:tim-visualization}),
and for exploring metrics in performance of testing \cite{abed2019measuring}.
Furthermore, the data has been used in hackathons in the AIDOaRt research project, \cite{eramo2021aidoart}, where also test case dependencies and anomaly detection of test automation performance was explored. This data was central in Strandberg's doctoral thesis \cite{strandberg2021automated}.
In short, students, researchers and practitioners in the field of software test automation, graphic design or artificial intelligence can benefit from the data. It can be used to evaluate algorithms, tools or visualizations for improved validity and generalizability.

\section{Objective}
The release of this data set is motivated by several factors:
\begin{enumerate}
\item Low harm on Westermo: the data set has already been collected so releasing it requires a low effort,
and the data is about ten years old, so any unexpected information leak that might stem from it is most likely not very harmful.
\item High value to research: realistic industry data is frequently requested by researchers, and the data has been used in closed contexts with great success (e.g.\ internal research, thesis students and hackathons).
\item Beneficence in general: Releasing the data might do good as new research, algorithms or tools could be valuable not only for researchers, but also for the general public, and for Westermo. 
\item Industry-academia relations: One often says that there is a distance between academia and industry, release of data could hopefully render researched solutions more realistic and would thereby lower thresholds for industry adoption of research artifacts, as well as simplify relations between academia and industry.
\end{enumerate}

\section{Data description}
There are two types of data in this data set: test results data and source code changes. The test results data is in one file, and source code change data is stored in more than a thousand text files. 
Both data types have been anonymized for ethical reasons.
Before going into details of the data format, this section first explains the anonymization process, and some aspects of the software development process in place at the time. (The industry context is further described in Section~\ref{experimental-design}.)

\subsection{Anonymization Process}
\label{anonymization-process}
A test case in the data set typically targets a communication protocol, e.g.\ firewall, RSTP, VRRP, DHCP, etc. The name and the path of the test cases are built up of one or more names of such protocols, and other words that give a very brief description of the test case, e.g., a test with the path \texttt{firewall/basic} could be a simple firewall test, whereas \texttt{dhcp/relay\_rstp\_firewall}, could be a test case that verifies the DHCP relay functionality in a network where the RSTP redundancy protocol is activated while using a firewall somehow. To protect Westermo's business interests, these keywords have been anonymized in the data, and replaced with items like \texttt{word7474} (these numbers were randomized). Using an ad-hoc stop-list, generic words like ``and,'' ``packet,'' etc.\ have been removed from the data set.

Furthermore, the exact dates for testing have been obfuscated, such that the first day of nightly testing has set to 1'st of February 1970. The dates for code changes have also been altered in the same manner. 

\subsection{WeOS Branches}
\label{weos-branches}
\begin{figure}
  \centering
  \includegraphics[width=0.95\linewidth]{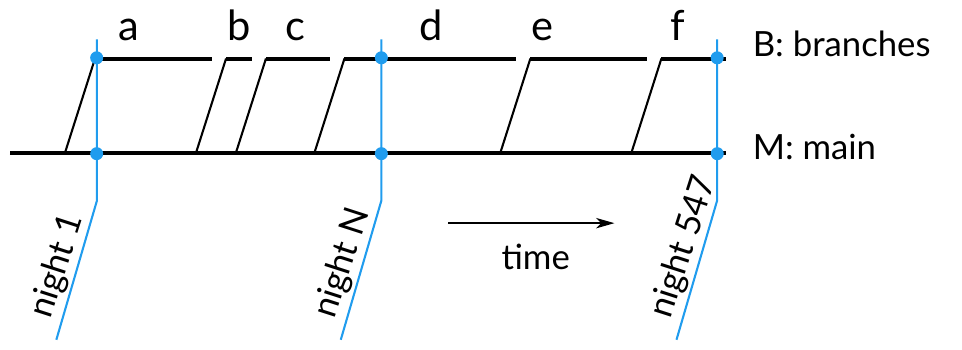}
  \caption{
  Illustration of the software process in place at the time of data collection.
  Each night, the main and branch tracks of the software were tested.
  }
  \label{fig:branches}
\end{figure}
At the time of data collection, the software under test was developed in a software process where two tracks of the software was active at the same time. In this data set we call these tracks main (M) and branch (B). The M track was very active, and most if not all code changes came in here. When a release for general availability was about to be made, a new B track was created. These were stable and the idea was that these should only receive stabilizing code changes, bug fixes, and updates in documentation, etc. Over time, new branches were generated, and this data set contains six of them, see Figure~\ref{fig:branches}. Each night, testing ran on both the M and the B release.

\subsection{Source code data}
\label{source-code-data}
The source code data is stored in more than one thousand small text files. Each file represents code changes from one day.
First in the \texttt{fawlty} folder, changes, day by day, of in the test framework can be found.
Second, in the \texttt{weos} folder changes in the main track of WeOS is found.
Third, in \texttt{weos-branches}, with one sub-folder per release branch (e.g. \texttt{weos-branches/a}), the code changes in branches are located.

Source code data is generated from the git log command, with a set of flags that also lists the name of the modified source code files. A source code change in the test framework may look something like this:

\begin{verbatim}
  Date: November 22, 2021
  Author: Per Erik Strandberg
  Message: Fixed a bug in the PoE wrapper and a test.

  Changed files:
   + src/lib/poe.py
   + src/test/poe/firewall_basic/main.py
\end{verbatim}

Now, if we assume that this is the only code change present on this day in the test framework, and that the keywords present in this code change are PoE (for Power over Ethernet), firewall, and basic; and that the anonymized keyword mappings
are 
PoE: word1001, 
firewall: word1002,
and 
basic: word1003; then
the file with code changes for this day would contain:
\begin{verbatim}
  word1001 3
  word1002 1
  word1003 1
\end{verbatim}

This translation comes from the fact that the word poe (word1001) occurs three times in the git log message, and that firewall and basic occur once each. (NB., these are not the actual word mappings).

\subsection{Test results data}
\begin{figure}
  \includegraphics[width=\linewidth]{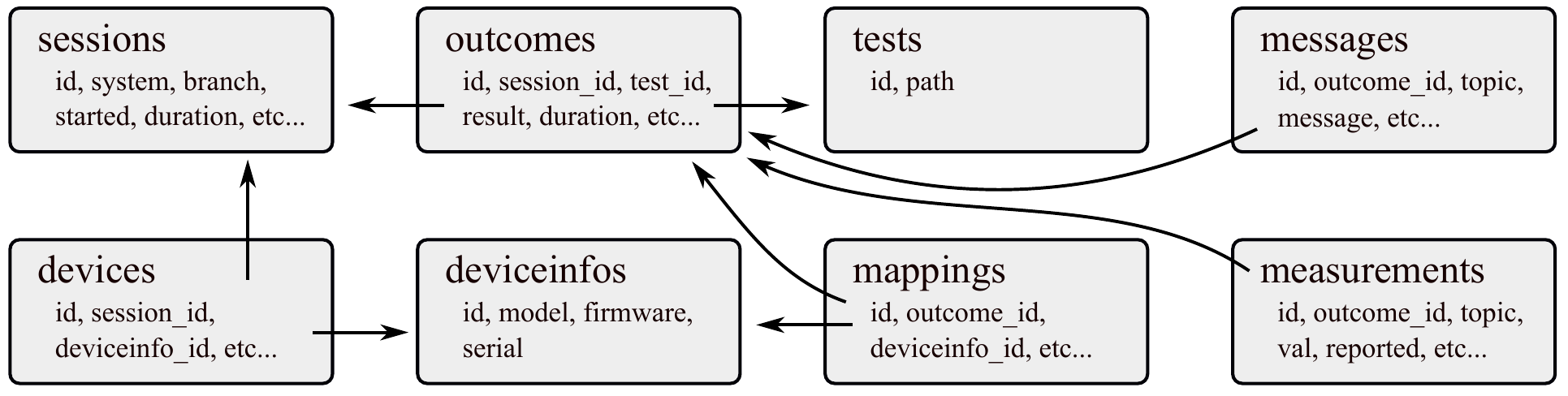}
  \caption{
  Illustration of the test results database schema.
  The first three are part of the data set.
  Figure originally from \cite{strandberg2022software}.
  }
  \label{fig:db-schema}
\end{figure}

The test results data represents an export from some of the tables in the Westermo test results database. The schema of this database is illustrated in Figure~\ref{fig:db-schema}. There are three major entities in the collected data:
\begin{itemize}
 \item sessions - a session is when we run a suite of tests on one
   test system with a certain software version and testware version
 \item tests - a test case is a python script that aims to verify a part of the software under test.
 \item outcomes - an outcome is the details from when we run a test in
   a session
\end{itemize}

Tables illustrated in Figure~\ref{fig:db-schema} that are not part of the data set contains details on the hardware used in the various test systems, as well as contents of unexpected error messages or qualitative metrics of e.g.\ protocols. 

The test results data is stored in a tab separated file, \texttt{test-results.csv},
with the following columns:
\begin{itemize}
 \item jid - jobid, together with the system name, the pair (jid, system) forms a unique key for a test session
 \item system - name of test system
 \item sessiontype - here, N represents the regular nightly suite. Other suites are: I which contains tests targeting software areas with reported issues, and Q for tests in quarantine, this could be new or suspicious tests.
 \item rel - rel is short for the software release being tested, as explained in Section~\ref{weos-branches}. M represents the main code branch in use at the time. B represents the other branch in use, which is a more stable release branch.
 \item sstarted - time of start for session
 \item sduration - duration in seconds of the session. If the session does not finish normally (e.g., when a human aborts it, or if the test framework would crash for some reasone), then duration is NULL. 
 \item fgstate - identifier of testware code state (a truncated git
   commit hash)
 \item oid - outcome id, unique for each execution of a test case
 \item tid - test id, unique per test case
 \item path - path to test case
 \item parameters - parameter settings of the test when it ran, if any
 \item result - the verdict of the test execution may be 0, 1, 2 or 3:
 \begin{itemize}
  \item 0 - pass: the test ran to completion and no problems were found.
  \item 1 - fail: the test ran to completion and at least one problem was found.
  \item 2 - invalid: the test did not run to completion, typically because of unexpected or unhandled states
  \item 3 - unmappable: the resources needed to run the test were not available
  \item In later use (not in this data set), the result 4 (unloadable) was used for tests that did not exist or had syntax errors, as well as the result 5 (skipped) for tests that were in queue when testing was stopped because time ran out.
  \end{itemize}
 \item ostarted - start of test case execution
 \item oduration - duration of test case execution
\end{itemize}

\section{Experimental design, materials and methods}
\label{experimental-design}
The source code data in this data set comes from the git source code repositories in use at Westermo at the time. Source code management tools such as git are standard and widely used among software engineering practitioners, not managing source code with such a tool is unimaginable for most software companies. Data was added in the repository as part of every day work. The source code data has been minimized and anonymized as described above (Sections~\ref{anonymization-process}~and~\ref{source-code-data}).

\begin{figure}
  \includegraphics[width=\linewidth]{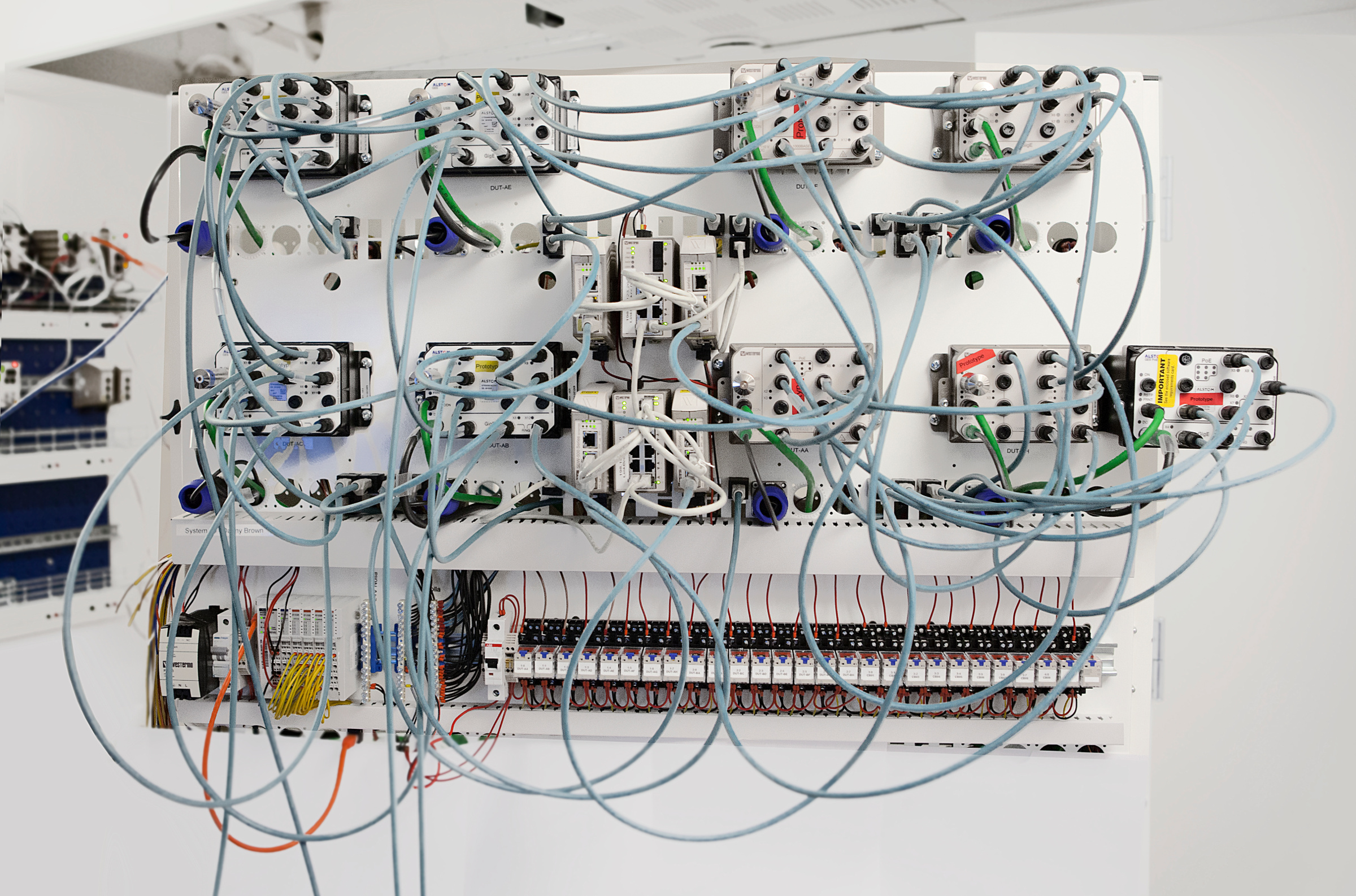}
  \caption{
  A Westermo test system with a network topology built up of switches, routers and other peripheral equipment.  
  }
  \label{fig:test-system}
\end{figure}

The test results data comes from running test cases in test systems such as the one illustrated in Figure~\ref{fig:test-system}. In particular, these test systems have real network topologies and are built up of real embedded systems -- Westermo products: switches and routers for industry needs (e.g., on board a train, for track-side signaling, in a power distribution site, or for industry automation in general). The test cases are implemented in code, and each test case can be thought of as having three steps:
\begin{itemize}
    \item First, each test case has a set of requirements and configuration needs that the test case has on the test system. One could imagine a test case that requires a very large network topology, or hardware products with serial ports. Not all test systems have large topologies, and not all hardware products have serial ports, so some test cases cannot have their requirements fulfilled by all test systems -- these get the result 3 (unmappable), as described above. 
    \item Second, for test cases can be mapped, the test framework now sets up the desired system state needed for the test. If a test case requires three devices, and a test system had four, then the last device has the ports deactivated in order to not disturb the network during the test phase. The first three has the required protocols activated, along with other needed configuration. If the test framework cannot get the test system to reach the desired state, then testing cannot start, and the test framework reports the result 2 (invalid).
    \item Third, in the test phase, the test framework goes through a number of stimuli, measurements and reconfigurations in order to verify that the software targeted by a test case works as expected. For a firewall test, this could mean sending traffic of various sorts through the network, capturing the traffic and analyzing that only the desired packets have gone through the firewall.
    If not all steps could be performed, then again, the result 2 (invalid) is reported.
    If the steps could run as expected but some issues in the software were found, then 1 (fail) is reported, otherwise 0 (pass).
\end{itemize}

\section{Ethics statements}
To protect Westermo and individuals that contributed to the data, we minimized and anonymized the data set. The public release of data has been approved by staff responsible for information security at Westermo Network Technologies AB at risk workshops during Q4 of 2022.

\section{CRediT author statement}
The author of this paper has contributed to 
Conceptualization,
Methodology,
Software,
Validation,
Investigation,
Resources,
Data Curation,
and 
Writing (both Original Draft and Review \& Editing).
Other Westermo staff has contributed with
Validation,
Resources,
and 
Supervision.

\section{Acknowledgments}
This work has been funded by 
Westermo Network Technologies AB,
the Swedish Knowledge Foundation through grants
20150277 (ITS ESS-H), and 20160139 (TESTMINE),
as well as 
the AIDOaRt project a European ECSEL Joint Undertaking (JU) under grant agreement No.\ 101007350.

\section{Declaration of interests}
The author of this paper is employed at Westermo Network Technologies AB.

\section{License}
This data set is licensed with the 
\href{https://creativecommons.org/licenses/by/4.0/}{Creative Commons Attribution 4.0 International}.

In short, you are free to: share, copy and redistribute the material; and to adapt, remix, transform, and build upon it for any purpose;
under the condition that you you give appropriate credit, and do not restrict others from doing anything the license permits. Read the license for details.

Suggested attribution:
Strandberg, P E. (2022). The Westermo test results data set. Retrieved from \href{https://github.com/westermo/test-results-dataset}{https://github.com/westermo/test-results-dataset}

\newpage
\bibliographystyle{abbrv}
\bibliography{references.bib} 

\end{document}